\def\L{{\cal L}}
\def\L{{\cal L}}

\def\A{{\cal A}}
\def\E{{\cal E}}
\def\I{{\cal I}}
\def\J{{\cal J}}
\documentstyle[12pt]{article}
\begin{document}
\begin{flushright}
DO-TH-94/27\\
(November 1994)
\end{flushright}
\vspace{3.8cm}
\centerline{\Large \bf {A general framework}}
\centerline{\Large \bf
 {with two Higgs doublets and suppressed FCNC~\footnote{partly based on
talk presented at Workshop on Electroweak Symmetry Breaking, Budapest,
July 11-13, 1994}}}
\vspace{0.8cm}
\begin{center}
{{\bf Gorazd Cveti\v c~\footnote{e-mail address:
cvetic@het.physik.uni-dortmund.de}} \\
{\it Inst.~f\"ur Physik, Universit\"at Dortmund, 44221 Dortmund, Germany}}
\end{center}
\vspace{1.5cm}
{\centerline {ABSTRACT}}



\vspace{0.2cm}

We investigated a general framework of the
Standard Model with two Higgs doublets (2HDSM) with suppressed
flavor-changing neutral currents (FCNC's). Loop-induced FCNC
(and CP-violating) effects, when confronted with experimental
constraints for the $K$-$\bar K$, $B$-$\bar B$ and $D$-$\bar D$
mixing and for the $(b \to s \gamma)$ decay,
provide us with constraints on the values of the dominant Yukawa
couplings of the charged Higgs, i.e., the values of the couplings
of $H^{\pm}$ to $t$ and $b$ quarks. Once these {\it low} energy
experimental data relevant for the mixings and for the mentioned decay,
as well as the theoretical uncertainties for the hadronic matrix
elements, are sufficiently reduced, such analyses may be able to rule
out the minimal SM and even certain special types of the 2HDSM's
(e.g., the popular ``type II'', and ``type I'' 2HDSM). In such a case,
a more general 2HDSM framework discussed here could still survive as
a viable framework. Eventual detection of the charged Higgs in high
energy experiments and the measurement of its mass would represent
important information that would additonally help to rule out or to
favor the various specific 2HDSM scenarios that are contained in the
discussed 2HDSM framework.



\newpage

\noindent{\large \bf {1.) Introduction}}

\vspace{0.2cm}

\small\normalsize

Experiments show that flavor-changing neutral currents (FCNC) are
in nature (at low energies) very suppressed
\begin{displaymath}
Br(K^0_L \to \mu^+\mu^-) \simeq (7.4 \pm 0.4) \cdot 10^{-9} \ , \qquad
|m_{B^0_H}-m_{B^0_L} | \simeq (3.36 \pm 0.40) \cdot 10^{-10}MeV \ ,
\end{displaymath}
\begin{displaymath}
 |m_{K_L}-m_{K_S}| \simeq (3.510 \pm 0.018) \cdot 10^{-12}MeV \ , \
 |m_{D^0_1}-m_{D^0_2}| < 1.32 \cdot 10^{-10}MeV \ , \
\end{displaymath}
\begin{displaymath}
 Br(b \to s \gamma) = (2.32 \pm 0.67) \cdot 10^{-4} \ \mbox{etc.}
\end{displaymath}
The various alternative models of electroweak interactions -
extensions of the minimal Standard Model (MSM) - must take into account
the FCNC suppression. The most conservative extensions of the MSM
are apparently the models with two Higgs doublets (2HDSM's).

The conditions for the one-loop FCNC suppression of contributions
coming from gauge boson loops, i.e., the allowed representations
of fermions, have been investigated some time
ago~\cite{paschos}. In addition, Glashow and Weinberg~\cite{paschos}
proposed for the Higgs sector the MSM (one Higgs doublet model)
and the ``type I'' and ``type II'' 2HDSM's. They proposed them as
models which,  in a ``natural'' way, have the zero value for the
flavor-changing renormalized Yukawa couplings in the neutral sector
(called from now on: FCN renormalized Yukawa couplings).
These two types of the 2HDSM's have
been widely discussed in the literature. Their Yukawa sector Lagrangians
with bare or renormalized quantities have the form: \\
\indent {\bf a)} ``type I''  $[$2HDSM(I)$]$ -
just one Higgs doublet (say, $H^{(1)}$) couples to fermions
\begin{equation}
{\L}^{(I)}_{\mbox{\scriptsize{Yukawa}}}  =           -  \sum_{i,j=1}^3 \lbrace
{\tilde D}_{ij}^{(q)}( \bar q^{(i)}_L H^{(1)} ) d^{(j)}_{R} +
{\tilde U}_{ij}^{(q)}( \bar q^{(i)}_L \tilde H^{(1)} ) u^{(j)}_{R} +
\mbox{h.c.}
\rbrace + \cdots \ ,
\label{2HD1}
\end{equation}
where the dots represent the terms for the leptons. The following
notations are used:
\begin{displaymath}
H  =  {H^{+} \choose H^0} \ , \ \tilde H  = i \tau_2 H^{\ast} \ ,
\quad q^{(i)} = {u^{(i)} \choose d^{(i)}} \ , \quad
q^{(1)} = {u \choose d} \ , \ q^{(2)} = {c \choose s} \ , \
q^{(3)} = {t \choose b} \ ,
\end{displaymath}
and similarly for the leptonic doublets $\ell^{(i)}$ containing Dirac
neutrinos and charged leptons.

This model is very closely related to the minimal SM (MSM), the
only difference in the Yukawa sector being the smaller vacuum expectation
value $\langle (H^0)^{(1)} \rangle_0 = v_1/\sqrt{2} < v/\sqrt{2}$ ($v \approx
246.22 \ GeV$), and hence the correspondingly larger Yukawa coupling
parameters.

{\bf b)} ``type II'' $[$2HDSM(II)$]$ -
one doublet ($H^{(1)}$) couples to the ``down-type'' right-handed
fermions $d_{R}$, $\ell_{R}^{-}$ and is responsible for the
``down-type'' masses; the other doublet
($H^{(2)}$) couples to the ``up-type'' fermions
$u_{R}$, $\nu_{R}$ and is responsible for their masses:
\begin{equation}
{\L}^{(II)}_{\mbox{\scriptsize{Yukawa}}}  =           -  \sum_{i,j=1}^3 \lbrace
{\tilde D}_{ij}^{(q)}( \bar q^{(i)}_L H^{(1)} ) d^{(j)}_{R} +
{\tilde U}_{ij}^{(q)}( \bar q^{(i)}_L \tilde H^{(2)} ) u^{(j)}_{R} +
\mbox{h.c.}
\rbrace + \cdots \ .
\label{2HD2}
\end{equation}
The mass matrices are proportional to the vacuum expectation values (VEV's)
of the Higgses:
$M_u^{(q,\ell)} = v_2 {\tilde U}^{(q,\ell)}/\sqrt{2}$,
$M_d^{(q,\ell)} = v_1 {\tilde D}^{(q,\ell)}/\sqrt{2}$ , where
\begin{displaymath}
\langle H^{(2)} \rangle_0 = \frac{e^{i\xi}}{\sqrt{2}} {0 \choose v_2} \ ,
\qquad \langle H^{(1)} \rangle_0 = \frac{1}{\sqrt{2}} {0 \choose v_1} \ ,
\qquad v_1^2+v_2^2=v^2 (\approx 246^2 GeV^2) \ .
\end{displaymath}
The phase $\xi$ between the two VEV's, if
it is nonzero, is responsible for CP violation in the scalar
and in the Yukawa sector.
The expressions are written in any $\mbox{SU(2)}_L$-basis, i.e., a basis
in which $q^{(i)}_L$ and $\ell^{(i)}_L$ are $\mbox{SU(2)}_L$-isodoublets.

\vspace{0.2cm}

Later on, extensions with more than one Higgs doublet other than
the 2HDSM(I) and (II) have been proposed. They usually satisfy
either one of the following conditions:
\begin{description}
\item[(a)] the renormalized FCN Yukawa couplings are {\it zero},
and the loop-induced FCNC phenomena are sufficiently
suppressed~\cite{strocchi}; these models possess an exact
family (``horizontal'') symmetry which ensures that
both the bare and the renormalized FCN Yukawa couplings
are zero simultaneously (in the formal limit of the infinite UV cut-off).
\item[(b)] FCN renormalized Yukawa couplings are nonzero, but small;
the suppression of these FCN Yukawa couplings
is brought about by an additional mechanism, e.g., by approximate
family symmetries~\cite{wu,cheng}; the FCN bare Yukawa couplings
are not necessarily small.
\end{description}
On the other hand,
we are going to consider several phenomenological consequences of
the following 2HDSM framework: within the framework, the renormalized
FCN Yukawa couplings are either zero or they are ``sufficiently''
suppressed. By ``sufficiently'' we mean that their suppression is
such that the leading 1-particle-irreducible (1PI) loop-induced
contributions to the considered FCNC processes and phenomena
(particularly the box diagrams for the $B$-$\bar B$ and
$D$-$\bar D$ mixing) dominate over the direct tree level contributions
of the renormalized FCN Yukawa couplings~\footnote{
The tree level contributions of the renormalized couplings are
in general tree level contributions of the bare couplings plus
(cut-off dependent parts of the)
1-particle-reducible (1PR) loop corrections calculated with these
bare couplings.}.
At the end of Section 4, we will estimate the upper bounds on the
relevant renormalized FCN Yukawa couplings satisfying the condition
of the ``sufficient'' suppression.
Physically, the considered framework includes models of type (a) and
a subset of the models of type (b). From the algebraic point of view,
the considered framework, although without imposed family symmetries, is
of type (a), because we will neglect the effects of the renormalized
FCN Yukawa couplings. In addition, for simplicity,
we assume that the CP-violating phase $\xi$ between the two VEV's
is so small that its effects will be neglected (by setting $\xi = 0$).
The CP violation then originates solely from the $\delta$ angle
of the Cabbibo-Kobayashi-Maskawa (CKM) mixing matrix.
The renormalized Yukawa interactions in this
framework, in any $\mbox{SU(2)}_L$-basis, have at first glance the most
general form
\begin{eqnarray}
{\L}_{\mbox{\scriptsize{Y.}}} & = & - \sum_{i,j=1}^3 {\Big \lbrace}
{\tilde D}_{ij}^{(1)}( \bar q^{(i)}_L H^{(1)} ) d^{(j)}_{R} +
{\tilde D}_{ij}^{(2)}( \bar q^{(i)}_L H^{(2)} ) d^{(j)}_{R} +
    \nonumber\\
& & + {\tilde U}_{ij}^{(1)}( \bar q^{(i)}_L \tilde H^{(1)} ) u^{(j)}_{R} +
{\tilde U}_{ij}^{(2)}( \bar q^{(i)}_L \tilde H^{(2)} ) u^{(j)}_{R} + h.c.
{\Big \rbrace}
+ \lbrace  \ \bar \ell\mbox{H}\ell\mbox{-terms} \ \rbrace \ .
\label{2HD30}
\end{eqnarray}
The quarks are here in an arbitrary $\mbox{SU(2)}_L$-basis (not in
the mass basis).
The renormalized $3 \times 3$-Yukawa matrices $\tilde D^{(1)}$,
$\tilde D^{(2)}$, $\tilde U^{(1)}$ and $\tilde U^{(2)}$
are in the present framework such that in the mass basis of quarks
they become all {\it simultaneously} diagonal, in
order to have zero (or: negligible) FCN Yukawa couplings. By that we mean
that ${\L}_{\mbox{\scriptsize{Y.}}}$ in the mass basis of quarks has the
following form:
\begin{eqnarray}
\lefteqn{
{\L}_{\mbox{\scriptsize{Y.}}}  =  - \sum_{i,j=1}^3  {\Big \lbrace}
\bar u^{(i)}_L \left[  H^{(1)+}  (V D^{(1)})_{ij} +
 H^{(2)+}  (V D^{(2)})_{ij} \right]
d^{(j)}_{R} +
\bar d^{(i)}_L \left[   H^{(1)0}  D^{(1)}_{ij} +
 H^{(2)0}  D^{(2)}_{ij} \right] d^{(j)}_{R}   }
 \nonumber\\
& &
+ \bar u^{(i)}_L \left[
\left( H^{(1)0} \right)^{\ast} U^{(1)}_{ij} +
  \left( H^{(2)0} \right)^{\ast} U^{(2)}_{ij} \right] u^{(j)}_{R} +
\bar d^{(i)}_L \left[
-  H^{(1)-}  (V^{\dagger} U^{(1)})_{ij}
-  H^{(2)-}  (V^{\dagger} U^{(2)})_{ij}
\right] u^{(j)}_{R}
\nonumber\\
& &
+ h.c. {\Big \rbrace} \ ,
\end{eqnarray}
\begin{displaymath}
\mbox{\noindent{ \ where:}} \qquad
H^{(k)}  = {  H^{(k)+}  \choose
              H^{(k)0}  } \ , \
           {\tilde H^{(k)}}  = i \tau_2 \left( H^{(k)} \right)^{\ast}
         = {  \left(H^{(k)0} \right)^{\ast} \choose
              - H^{(k)-} } \ , \ (k=1,2) \ .
\label{2HD3}
\end{displaymath}
In formula (\ref{2HD3}), the quarks are in the mass basis, $V$ is the usual
(complex) CKM matrix, and $U^{(1)}$, $U^{(2)}$, $D^{(1)}$ and
$D^{(2)}$ are the diagonal renormalized Yukawa matrices (in the
mass basis), i.e., the FCN Yukawa couplings are all zero.
Furthermore, these matrices are all real, because we assume that
all the CP violation is of the CKM-type only ($\xi = 0$) and
can therefore be presented with a single $\delta$-angle in the
complex CKM matrix. We note that ``type I'' and ``type II'' models
(eqs.~(\ref{2HD1}) and (\ref{2HD2})) are special cases (subsets)
of this framework. In (\ref{2HD3}) we omitted the leptonic sector.

There may be objections against such a 2HDSM framework, on the grounds
that, unlike the 2HDSM(I) and (II), it has no discrete or continuous
family (``horizontal'') symmetries. These family symmetries
would impose in a ``natural'' way the value zero on the bare FCN Yukawa
couplings, and would keep these Yukawa couplings at the value zero
even when they are formally renormalized from $\Lambda = \infty$
to low energies~\footnote{
In the 2HDSM(II), the family symmetry in ${\L}_{\mbox{\scriptsize{Y.}}}$
is of U(1)-type:
$d^{(j)}_{R} \rightarrow \mbox{e}^{i \alpha} d^{(j)}_{R}$,
$ \ H^{(1)} \rightarrow \mbox{e}^{-i \alpha} H^{(1)}$ (j=1,2,3),
the other fields remaining unchanged. This symmetry ensures that,
in the course of renormalization, no loop-induced ($\ln \Lambda$
cut-off dependent) Yukawa couplings other
than those of the form (\ref{2HD2}) can appear.
In the 2HDSM(I), the family symmetry is similar:
$d^{(j)}_R \rightarrow \mbox{e}^{i \alpha} d^{(j)}_R$,
$ \ u^{(j)}_R \rightarrow \mbox{e}^{-i \alpha} u^{(j)}_R$,
$ \ H^{(1)} \rightarrow e^{-i \alpha} H^{(1)}$. }.
It is true that the renormalized
FCN Yukawa couplings in (\ref{2HD3}) acquire the value zero
(in the mass basis) not in a ``natural'' way, but only as
a consequence of an algebraic condition on the
renormalized Yukawa couplings - the condition that $U^{(1)}$, $U^{(2)}$,
$D^{(1)}$ and $D^{(2)}$ all be diagonal.
In general, such a condition leads to
the following behavior of the FCN Yukawa couplings: at low
energies of probes, i.e., when they are renormalized, they are
(negligibly) small; only as the energy of probes is increased,
they may in general grow, as dictated by the renormalization group
equations (RGE's). In the formal limit of $E_{\mbox{\scriptsize{probes}}}$
($= \Lambda$)
$= \infty$, the resulting bare FCN Yukawa couplings may in general
diverge. In view of this, the objection against the framework can be
countered in at least two different ways:
\begin{itemize}
\item Since the FCNC suppression is an essentially low energy phenomenon
($E_{\mbox{\scriptsize{probes}}} \stackrel{<}{\sim} 100 \ GeV$), there is
no absolutely
compelling reason for it to remain in force when the energy of probes
is increased beyond the present experimentally accessible regions and
relevant phenomena and processes are investigated at such an energy.
Then, the (negligibly) small values of the FCN Yukawa couplings
at low energies could be ensured by some as yet unknown mechanism
(which is definitely other than exact family symmetries), e.g.,
by an approximate family symmetry (cf.~Refs.~\cite{wu,cheng}),
or could be simply accidental.
\item Even if we assume the ``naturality'' of the FCNC's,
i.e., that the suppression of the FCN Yukawa couplings to zero
remains in force even at increasing $E_{\mbox{\scriptsize{probes}}}$
because ensured by certain assumed exact family symmetries,
we can regard the considered framework (\ref{2HD3}) as a framework
containing the union of all such ``natural'' 2HDSM's. Even in such
a case, the phenomenological analysis of the framework (\ref{2HD3}) would
carry relevance, because such an analysis does not investigate theoretical
constraints of a specific model, but only phenomenological constraints
in a rather broad framework.
\end{itemize}
In Section 2 we discuss the charged sector of the Yukawa couplings in
the proposed 2HDSM framework. In Section 3 we investigate phenomenological
constraints on this sector which are imposed by the $K$-$\bar K$ and
$B$-$\bar B$ mixing data. In Sections 4 and 5 we investigate the constraints
on this sector which come from the $D$-$\bar D$ mixing and from the
$(b \to s \gamma)$ decay data, respectively. The conclusions are
summarized in Section 6.

\vspace{0.5cm}

\noindent{\large \bf {2.) The Yukawa sector of
the charged Higgs in a general 2HDSM
framework}}

\vspace{0.3cm}

The charged-current part of the quarks corresponding to the Lagrangian
(\ref{2HD3}) can be deduced in a straightforward manner
\begin{equation}
{\L}^{cc}_{\mbox{\scriptsize{Y.}}} = H^{(+)} \lbrack - \overline{u_L}V D d_R +
\overline{u_R} U
V d_L \rbrack + H^{(-)} \lbrack -\overline{d_R}D V^{\dagger} u_L
+ \overline{d_L} V^{\dagger} U u_R \rbrack \ .
\label{ccpart}
\end{equation}
This expression is in the unitary gauge and in the physical bases
of the quarks {\it and} of the charged Higgses $H^{(\pm)}$. $V$ is the
usual (complex) CKM matrix, $u^T = (u,c,t)$, $d^T = (d,s,b)$;
$U$ and $D$ are specific linear combinations of the real diagonal
Yukawa matrices $U^{(j)}$ and $D^{(j)}$
\begin{equation}
U  =  - \sin \beta U^{(1)} + \cos \beta U^{(2)} \ ,
\qquad
D  =  - \sin \beta D^{(1)} + \cos \beta D^{(2)} \ ,
\label{Yukawas}
\end{equation}
where $\tan \beta$ is the ratio of the absolute values of the
VEV's ($\tan \beta = v_2/v_1$), and the mass basis is used. $U$ and $D$
are diagonal and real ($\xi = 0$).
On the other hand, the (diagonal) mass matrices of quarks are obtained from
the corresponding orthonormal linear combinations
\begin{equation}
\frac{\sqrt{2}}{v}M_u  =
 \cos \beta U^{(1)} + \sin \beta U^{(2)} \ ,
\qquad
\frac{\sqrt{2}}{v}M_d  =
 \cos \beta D^{(1)} + \sin \beta D^{(2)} \ .
\label{masses}
\end{equation}
Therefore, if we assume that no peculiar cancelations occur,
we have the following hierarchy:
\begin{equation}
\frac{U_{ii}}{U_{jj}} \sim \frac{m_u^{(i)}}{m_u^{(j)}}, \qquad
\frac{D_{ii}}{D_{jj}} \sim \frac{m_d^{(i)}}{m_d^{(j)}} \qquad
(i,j=1,2,3) \ .
\label{hierarchy}
\end{equation}
In such a case, the coupling
$U_{33}$ of the charged Higgs $H^{(+)}$ to $\overline{t_R} b_L$
and the coupling $(-D_{33})$ of $H^{(+)}$ to $\overline{t_L}b_R$
are the two dominant Yukawa couplings in the charged sector.
We will assume in the rest of the presentation that this is the
case, i.e., that the hierarchy (\ref{hierarchy}) holds.
Stated otherwise, the sector of the Yukawa couplings of charged
Higgs to quarks provides us, within the present framework, with only
two additional real ``normalized'' parameters
\begin{equation}
X^{(U)}(E)= \frac{U_{33}(E) v}{m_t(E) \sqrt{2}} \ , \qquad
X^{(D)}(E)= - \frac{D_{33}(E) v}{m_b(E) \sqrt{2}} \ ,
\label{defp}
\end{equation}
where $E$ is a typical energy of probes which depends on the process
considered~\footnote{
Formally, $E$ is the finite upper energy cut-off used in the renormalized
Lagrangians (\ref{2HD30})-(\ref{ccpart}).
Later we will argue that this energy $E$, for the quantities
$U_{33}(E)$ and $D_{33}(E)$ (or $X^{(U)}(E)$ and $X^{(D)}(E)$)
appearing in one-loop formulas for specific processes, is approximately
equal to the higher of the two momenta (or masses) of the two
quark legs at the vertex $U_{33}$ or $D_{33}$. This
energy turns out to be $E \simeq m_t$ for the $B$-$\bar B$ and
$K$-$\bar{K}$ mixing and for the ($b \to s \gamma$) decay,
and $E \simeq m_b$ for the $D$-$\bar{D}$ mixing.}.
2HDSM(I) and (II) [cf.~Eqs.~(\ref{2HD1}),(\ref{2HD2})]
are just two special cases of the present
framework, and $X^{(U)}$ and $X^{(D)}$ are in these cases
independent of $E$:
\begin{equation}
\mbox{2HDSM(I):} \qquad
U^{(1)} = D^{(1)} = 0 \, \qquad X^{(U)}(E) = - X^{(D)}(E) = \cot \beta \ .
\label{typeI}
\end{equation}
\begin{equation}
\mbox{2HDSM(II):} \qquad
U^{(1)} = D^{(2)} = 0 \ , \qquad X^{(U)}(E) = 1/X^{(D)}(E) = \cot \beta \ .
\label{typeII}
\end{equation}
Therefore, if abandoning the present general framework in favor
of the 2HDSM(I) or (II), the two parameters $X^{(U)}$ and $X^{(D)}$
would not be free any more, but would be completely fixed by
the VEV's (i.e., by $\tan \beta$) and would be interrelared.

In the analysis below, we will search for constraints on these two
additional parameters $X^{(U)}$ and $X^{(D)}$. The constraints
will be dictated by the experimental data on various meson-antimeson
mixings and the $(b \to s \gamma)$ decay. In principle, these essentially
{\it low}
energy phenomena alone could at some future point, when experimental and
theoretical uncertainties are reduced, give us a possibility to rule out
the MSM, or even to rule out 2HDSM(II) and (I). In the latter case,
the constraint $X^{(U)} \cdot X^{(D)} = 1$ and/or the constraint
$X^{(U)}=-X^{(D)}$ would have to be abandoned in favor of a
more general framework, e.g.~the 2HDSM framework discussed here.
For the special cases 2HDSM(II) and (I), phenomenological constraints
have been investigated by several authors~\cite{gg}. Here they will
be studied within the presented more general framework. However, before
doing this, we note that certain constraints on $X^{(U)}$ and
$X^{(D)}$ can be immediately obtained by demanding that the theory
behave perturbatively. This demand can be implemented approximately,
by requiring that the relevant Yukawa couplings $U_{33}(1 \pm \gamma_5)/2$
and $D_{33}(1 \pm \gamma_5)/2$ of the charged Higgs to $b$ and $t$
quarks [cf.~Eq.~(\ref{ccpart})] not exceed the QCD coupling
$g_s^2(M_W^2)$ ($ = 4 \pi \alpha_s(M_W^2)$ $\simeq 1.5$)
\small
\begin{equation}
\frac{D_{33}(m_b)}{2}, \ \frac{D_{33}(m_t)}{2}, \ \frac{U_{33}(m_t)}{2}
 \stackrel{<}{\sim}  1.5 \ \Rightarrow
|X^{(D)}(m_b)| \stackrel{<}{\sim} 120  , \
|X^{(D)}(m_t)| \stackrel{<}{\sim} 190  , \
|X^{(U)}(m_t)| \stackrel{<}{\sim}   3.0 \ .
\label{pert}
\end{equation}
\normalsize


\noindent{\large \bf {3.) $B$-$\bar B$ and $K$-$\bar{K}$ mixing
in a general 2HDSM
framework}}

\vspace{0.3cm}

Since the $H^{(\pm)}$-exchanges influence the short distance
contributions to the $B^0_d$-$\bar{B^0_d}$ and $K^0$-$\bar{K^0}$
mixing, the experimental values of
$|m_{B^0_H}-m_{B^0_L}|$, and of the $K^0$-$\bar{K^0}$
CP-violating parameter $\varepsilon_K$ would provide us
with restrictions on the (dominant) $X^{(U)}$ coupling
strength of $H^{(\pm)}$ to quarks.
These parameters are dominated by short distance physics,
because it is mostly the very heavy top quark that dominates
over the other quark contributions in the relevant electroweak loop
diagrams. The dominant one-loop
electroweak diagrams contributing to these
mass differences are the W-W, H-W and H-H exchange box diagrams
of Fig.~1. The resulting effective four-fermion couplings are
\begin{eqnarray}
{\L}_{eff}  (= {\L}_{eff}^{WW} + {\L}_{eff}^{HW} + {\L}_{eff}^{HH})
 & \simeq & {\A}^K \left[ \overline{d(x)^a} \gamma^{\mu}
(\frac{1-\gamma_5}{2})s(x)^a \right]^2 \quad
(\mbox{for }  \bar{K^0} \to K^0)
\nonumber\\
&  \simeq & {\A}^B \left[ \overline{b(x)^a} \gamma^{\mu}
(\frac{1-\gamma_5}{2})d(x)^a \right]^2
\quad (\mbox{for } B_d^0 \to \bar{B_d^0}) \ ,
\label{leff}
\end{eqnarray}
where ${\A}^K$ and ${\A}^B$ are the corresponding box amplitudes from
Fig.~1
\begin{displaymath}
\qquad {\A}  =  {\A}_{WW} + {\A}_{HW} + {\A}_{HH} \ .
\end{displaymath}
The general formulas for these electroweak amplitudes, within the present
framework, can be calculated in a straightforward way, if we ignore
the masses and momenta of the external quark legs
\begin{equation}
{\A}_{WW}  =  \frac{G_F^2 M_W^2}{4 \pi^2} \sum_{k,n=2}^3
\zeta_k \zeta_n {\E} \left( x_k,x_n \right) \ , \qquad
\A_{HH}  =  - \frac{1}{128 \pi^2 M_{H^{\pm}}^2} \sum_{k,n=1}^3
\zeta_k \zeta_n U_{kk}^2 U_{nn}^2 {\I} \left( y_k,y_n \right) \ .
\label{ahh}
\end{equation}
\begin{equation}
{\A}_{HW}  =  -\frac{G_F}{4 \sqrt{2} \pi^2} \sum_{k,n=1}^3 \zeta_k \zeta_n
\sqrt{x_k x_n} U_{kk} U_{nn} {\J} \left( x_k,x_n,x_h \right) \ ,
\label{ahw}
\end{equation}
We denoted here $x_k=(m_k^{(u)}/M_W)^2$, $y_k=(m_k^{(u)}/M_{H^{\pm}})^2$
($k=1,2,3$), $x_h=(M_{H^{\pm}}/M_W)^2$; $\zeta_k$ are the CKM-mixing
factors
\begin{equation}
\zeta_3 = \zeta_t  =  V^{\ast}_{td} V_{ts} \
(\mbox{for } \bar{K^0} \to K^0) \ ,
 \qquad
\zeta_3 = \zeta_t = V_{td}V^{\ast}_{tb} \ (\mbox{for } B^0 \to \bar{B^0}) \ ,
\end{equation}
and analogously for $\zeta_2$ and $\zeta_1$. The integrals
$\E$, $\J$ and $\I$ are dimensionless and tame functions of the masses of
internal particles
\begin{eqnarray}
{\J}(x_k,x_n;x_h)&=& \int_0^{\infty} \frac{dz z (1+ z/4)}{(z+1)(z+x_h)
 (z+x_k)(z+x_n)} \ ,
\nonumber\\
{\I}(y_k,y_n)&= &\int_0^{\infty} \frac{dz z^2}{(z+1)^2 (z+y_k) (z+y_n)} \ .
\label{inami}
\end{eqnarray}
${\E(x_k,x_n)}$ is a well known Inami-Lim function~\cite{inami}.
The integration variable $z$ in the above expressions for ${\J}$ and ${\I}$
is equal to ${\bar p}^2/M_W^2$ and ${\bar p}^2/M_{H^{\pm}}^2$, respectively,
${\bar p}$ being the Euclidean version of the 4-momentum of the loop.
In view of the previous discussion, we will neglect in the above
expressions for ${\A}_{HW}$ and ${\A}_{HH}$ all the terms containing
$U_{11}$ and $U_{22}$. Therefore, ${\A}_{HW}$ and ${\A}_{HH}$
contain each just one term, namely the term proportional to
$\zeta_t^2 x_t U_{33}^2$ and to $\zeta_t^2 U_{33}^4$, respectively.
In the above expression for ${\L}_{eff}^{HH}$, we ignored several
other induced four-fermion terms, e.g.~terms of the type
$[\overline{b(x)^a} \left( \frac{1-\gamma_5}{2} \right) d(x)^a]^2$.
It turns out that the dimensionless integrals appearing in the
amplitudes at such terms are roughly an order of magnitude smaller
than the integrals ${\I}$ and ${\J}$ in eqs.~(\ref{inami}). Furthermore,
the dominant coupling strengths at such terms are proportional to
$\zeta_t^2 U_{33}^2 D_{22}^2$ for $K^0$-$\bar{K^0}$, and to
$\zeta_t^2 U_{33}^2 D_{33}^2$ for $B_d^0$-${\bar{B_d^0}}$.
Neglecting such terms is justifiable if
$|D_{33}| \stackrel{<}{\sim} |U_{33}|$ (at $E \simeq m_t$),
i.e., if $|X^{D}| \stackrel{<}{\sim} 60 \cdot |X^{(U)}|$ (in 2HDSM(II),
this condition reads: $\tan \beta \stackrel{<}{\sim} 8$)~\footnote{
We assumed here $m_t^{\mbox{\scriptsize{phy}}}=175 GeV$ and
$m_b^{\mbox{\scriptsize{phy}}}=4.9 GeV$,
which corresponds to $m_t(m_t) \simeq 167 GeV$ and $m_b(m_t) \simeq
2.77 GeV$.}.
We will see
in the analysis of the $(b \to s \gamma)$ decay that this condition
is satisfied in most of the region in the plane $X^{(D)}$ vs $X^{(U)}$
allowed by this decay.

The next step is to include in the above formulas the QCD corrections
in the leading logarithmic approximation. Many authors have investigated
these corrections for the box exchange diagrams in the MSM (\cite{vainshtein,
datta,gilman,buras} and references therein)~\footnote{
The authors of~\cite{buras} have calculated even the next-to-leading
order QCD corrections in an apparently consistent way.}. Applying the
approach of~\cite{datta} to the W-H and H-H box exchange diagrams (which
contain two top quark propagators), we arrive at the following
leading QCD correction factors to the integrands of (\ref{inami}):
\begin{description}
\item[(a):] The leading-log summation for the exchange of gluons
between the external quark legs of the same flavor (Fig.~2a)
yields the factor
\begin{equation}
\left[ \frac{\alpha_s(p^2)}{\alpha_s(\mu^2)} \right]^{
{\mbox{\footnotesize{2/b}}}_{\mbox{\scriptsize{nf}}}}
\qquad ({\mbox{b}}_{\mbox{nf}}=11-\frac{2}{3} {\mbox n}_{
\mbox{\footnotesize{f}} }) \ .
\label{qcda}
\end{equation}
Here, $p^2$ is the 4-momentum of the electroweak loop
(to be integrated over), ${\mbox n}_{\mbox{\footnotesize{f}} }$
is the effective number of flavors
at this energy (${\mbox n}_{\mbox{\footnotesize{f}} } \simeq 5$
for $|p^2| < m_t^2$, ${\mbox n}_{\mbox{\footnotesize{f}} } \simeq 6$
for $|p^2| > m_t^2$), and $\mu$ is the lower (infrared) energy
cut-off which is roughly to be identified with the momenta of
the quark constituents of the meson. We will take
$\mu = m_b \approx 4.9 GeV$ ($\Rightarrow \alpha_s(\mu^2) \approx 0.21$)
for $B_d^0$-${\bar{B_d^0}}$ mixing, as suggested by
Buras {\it et al.}~\cite{buras},
and $\alpha_s(\mu^2) \approx 1$ for $K^0$-$\bar{K^0}$ mixing, as
suggested by several authors~\cite{vainshtein, datta}.
\item[(b):] The effects of the ``running'' top mass $m_t(p^2)$ and
the ``running'' Yukawa couplings $U_{33}(p^2)$ (cf.~Figs.~2b,c)
yield the factor
\begin{equation}
\left[ \frac{\alpha_s(p^2)}{\alpha_s(m_t^2)} \right]^{
{\mbox{\footnotesize{16/b}}}_{\mbox{\scriptsize{nf}}}}.
\label{qcdb}
\end{equation}
\item[(c):] The effects of the gluon exchange between an outer leg
and the internal (top quark) propagator (cf.~Fig.~2d) do not contribute
appreciably, because the top quark is so heavy. We can intuitively
explain this by imagining that the top quark is so heavy as to have no
propagator, i.e., the propagator shrinks to a point, and the gluon
has ``no place'' to land on it. This argument holds as long as
$M_{H^{\pm}}$ is not exceedingly high (i.e., as long as $M_{H^{\pm}} \sim
m_t$).
\end{description}
It turns out that the factor in (b) is rather close to unity
for most relevant momenta (as long as $M_{H^{\pm}} \sim m_t$),
and that therefore the factor in (a) is the crucial one numerically.
The QCD-corrected integrals of (\ref{inami}) are
\begin{equation}
{\J}(x_k,x_n;x_h)=
\int_0^{\infty} \frac{dz z (1+ z/4)}{(z+1)(z+x_h)(z+x_k)(z+x_n)}
\cdot
\left[ \frac{\alpha_s(z M_W^2)}{\alpha_s(\mu^2)} \right]^{
\frac{ \mbox{\footnotesize{2}} }{ {\mbox{\footnotesize{b}}}_{
\mbox{\scriptsize{nf}}} } }
\cdot
\left[ \frac{\alpha_s(z M_W^2)}{\alpha_s(m_t^2)} \right]^{
\frac{ \mbox{\footnotesize{16}} }{ {\mbox{\footnotesize{b}}}_{
\mbox{\scriptsize{nf}}} } }
\label{Jqcd}
\end{equation}
\begin{equation}
{\I}(y_k,y_n) =
\int_0^{\infty} \frac{dz z^2}{(z+1)^2 (z+y_k) (z+y_n)}
\cdot
\left[ \frac{\alpha_s(z M_{H^{\pm}}^2)}{\alpha_s(\mu^2)}
   \right]^{
\frac{ \mbox{\footnotesize{2}} }{ {\mbox{\footnotesize{b}}}_{
\mbox{\scriptsize{nf}}} } }
\cdot
\left[ \frac{\alpha_s(z M_{H^{\pm}}^2)}{\alpha_s(m_t^2)}
  \right]^{
\frac{ \mbox{\footnotesize{16}} }{ {\mbox{\footnotesize{b}}}_{
\mbox{\scriptsize{nf}}} } }
\label{Iqcd}
\end{equation}
We calculated numerically these integrals, taking for $\alpha_s(p^2)$
the two-loop solution with $\alpha_s(M_Z^2)=0.117$. The results of the
analysis for $B_d^0$-${\bar{B_d^0}}$ and $K^0$-$\bar{K^0}$ mixing
depend only very weakly on the precise value of $\alpha_s(M_Z^2)$.
Furthermore, we included the leading (known) QCD corrections of the
MSM case, i.e., to the $W$-$W$ exchange box
diagrams~\cite{vainshtein}-\cite{buras}
($W$ stands here for the {\it physical} gauge boson $W$), by making the
following replacements in the amplitude
${\A}^K_{WW}$: $\zeta_c^2 \mapsto 0.81 \zeta_c^2$,
$\ \zeta_t \zeta_c \mapsto 0.37 \zeta_t \zeta_c$;
$\ \zeta_t^2 \mapsto 0.57 \zeta_t^2$,
and in ${\A}^B_{WW}$:
$\zeta_t^2 \mapsto 0.86 \zeta_t^2$.
The replacements for the $\zeta_t^2$-terms in the cases
of both mixings were taken from Ref.~\cite{buras}, where they
are substantiated by careful investigation of the next-to-leading
QCD corrections in the MSM. We note that it is the $\zeta_t^2$-terms
which contribute the most to the short distance physics in the
case of the $K$-$\bar K$ mixing, and almost exclusively in the case
of the $B$-$\bar B$ mixing.

The correction factor (\ref{qcdb}), although numerically not
important in the integrals (\ref{Jqcd}) and (\ref{Iqcd}),
suggests that the relevant scale
$E$ ($=\sqrt{p^2}$) to be used in the vertex coupling $U_{33}(E)$,
as well as in the mass $m_t(E)$ [in $x_t= (m_t(E)/M_W)^2$ and
$y_t=(m_t(E)/M_{H^{\pm}})^2$ in Eqs.~(\ref{Jqcd}) and (\ref{Iqcd})],
is $E \simeq m_t$. More generally, the scale $E$ used in
the ``running'' vertex couplings $U_{33}(E)$ and $D_{33}(E)$ appearing
in the one-loop (electroweak) formulas is approximately equal to the
mass $m_q$ of the heavier of the two quarks attached to the considered
vertex. This is connected to the fact that the largest contribution
to the electroweak one-loop integral is for the internal loop momenta
in the region $p^2 \sim m_q^2$. This argument appears to be
supported by the works~\cite{vainshtein, datta} (these works
suggest the formula (\ref{qcdb})) and by the work of Buras {\it
et al.}~\cite{buras}~\footnote{
The authors of~\cite{buras} also argue that the mass $m_t(E)$ in the internal
top quark propagators of the electroweak loop [i.e., in
$x_t(E)$ and $y_t(E)$ in integrals (\ref{Jqcd}) and (\ref{Iqcd})]
should be taken at $E \approx m_t$.}.
By adjusting the ``running'' energy of probes $E$ in the electroweak
formulas in this way appears to take
into account some QCD-correction effects which turn out to be quite
important in some processes. For example, in the ($b \to s \gamma$)
decay the electroweak formulas contain both $U_{33}(E)$ and
$D_{33}(E)$ Yukawa couplings, or equivalently: $X^{(U)}(E)$ and $X^{(D)}(E)$
(see Section 5). Since the relevant
electroweak loops are dominated by the top quark contributions
(cf.~Fig.~4b), we have $E \simeq m_t$. When normalizing $X^{(D)}$
according to (\ref{defp}), we therefore use $m_b(m_t)$ ($\simeq
2.8 \ GeV$) and not $m_b(m_b)$ ($\simeq 4.4 \ GeV$). On the
other hand, the charged Higgs contributions to the $D^0$-$\bar{D^0}$
mixing are dominated by the box diagrams
containing the bottom quarks. Therefore, the formulas there,
containing $X^{(D)}(E)$, should have $E \simeq m_b$.
All in all, the $B$-$\bar B$ and $K$-$\bar{K}$ mixing will result
in constraints on $X^{(U)}(m_t)$, the $D$-$\bar{D}$ mixing in
constraints on $X^{(D)}(m_b)$, and the ($b \to s \gamma$) decay
in constraints on $X^{(U)}(m_t)$ and $X^{(D)}(m_t)$.

Relations of the resulting amplitudes ${\A}^K$ and ${\A}^B$
to the experimental inputs of K and B physics are well known
\begin{eqnarray}
\sqrt{2} (\triangle M_{K^0_L-K^0_S}) (|\varepsilon_K| +
0.05 \frac{\varepsilon^{\prime}_K}{\varepsilon_K}) & \simeq &
- \frac{1}{2 M_{K^0}}
 \mbox{Im} \langle K^0 | \L_{eff}(x=0) | \bar{K^0} \rangle
\nonumber\\
& = & - \frac{1}{2 M_{K^0}} \mbox{Im} \left( {\A}^K \right) \cdot
  \langle K^0 | \left[ \overline{d^a} \gamma^{\mu}
(\frac{1-\gamma_5}{2})s^a \right]^2 | \bar{K^0} \rangle \ ,
\label{ekrel}
\end{eqnarray}
\begin{equation}
M_{B^0} |\triangle M_{B^0_H-\bar{B^0_L}}| \simeq
| \langle \bar{B^0_d} | \L_{eff}(x=0) | B^0_d \rangle | =
|{\A}^B| \cdot \langle \overline{B^0_d} | \left[ \bar{b^a} \gamma^{\mu}
(\frac{1-\gamma_5}{2})d^a \right]^2 | B^0_d \rangle  \ ,
\label{mbrel}
\end{equation}
where we adopt the normalization conventions: $\langle P^0|P^0 \rangle =
Vol \cdot 2 M_{P^0}$ ($P^0=K^0, B^0, \ldots$; $Vol$ is the
3-dimensional volume of the space). Perturbative (short distance)
contributions in the above relations are represented
by the amplitudes $\mbox{Im}({\A}^K)$, ${\A}^B$. On the other
hand, the hadronic matrix elements $\langle K^0 | \cdots | \bar{K^0} \rangle$
and $\langle \bar{B^0_d} | \cdots | B^0_d \rangle$
represent the low energy (non-perturbative) effects. Experiments provide us
with the following values for the $\triangle M$'s and the CP-violating
parameters $\varepsilon_K$ and $\varepsilon^{\prime}_K$~\cite{particle}:
\small
\begin{equation}
\triangle M_{K^0_L-K^0_S} =
3.510 \cdot 10^{-15} GeV (1 \pm 5.1 \cdot 10^{-3}) \ ,
\qquad
\triangle M_{B_H^0-B_L^0} = 3.36 \cdot 10^{-13} GeV (1 \pm 0.12) \ ,
\label{deltamb}
\end{equation}
\begin{equation}
|\varepsilon_K| = 2.26 \cdot 10^{-3} (1 \pm 0.037) \ , \qquad
\frac{\varepsilon^{\prime}_K}{\varepsilon_K} \simeq 1.5 \cdot 10^{-3} \ ,
\qquad M_{B_d^0} = 5.279 \ GeV \ , \ M_{K^0}=0.4977 \ GeV \ .
\label{delteps}
\end{equation}
\normalsize
Among these parameters, only $\triangle M_{B_H^0-B_L^0}$ has
an appreciable experimental uncertainty (12 percent).
For the hadronic matrix elements, we have the following theoretical
uncertainties:
\begin{equation}
\langle K^0 | \left[ \bar{d^a} \gamma^{\mu} (\frac{1-\gamma_5}{2})s^a
\right]^2 |
\bar{K^0} \rangle  =  \frac{2}{3} F^2_K B_K M_{K^0}^2 \ , \qquad
0.5 \stackrel{<}{\sim} B_K \stackrel{<}{\sim} 1.0 \ , \
(F_K=160 MeV) \ .
\label{bkgen}
\end{equation}
\begin{equation}
\langle \bar{B^0_d} | \left[ \bar{b^a} \gamma^{\mu}(\frac{1-\gamma_5}{2})
d^a \right]^2 | B^0_d \rangle  =  \frac{2}{3} F_B^2 B_B M_{B^0}^2 \ , \qquad
0.12 GeV \stackrel{<}{\sim} F_B \sqrt{B_B} \stackrel{<}{\sim} 0.25 GeV \ .
\label{fbprgen}
\end{equation}
We note that the QCD-sum rules and lattice calculations prefer the
following values of $B_K$ and $F_B \sqrt{B_B}$:
\begin{equation}
0.6 \stackrel{<}{\sim} B_K \stackrel{<}{\sim} 0.9 \ , \qquad
0.17 GeV \stackrel{<}{\sim} F_B \sqrt{B_B} \stackrel{<}{\sim} 0.22 GeV \ .
\label{bkrestr}
\end{equation}
In addition, we have many uncertainties also in the CKM matrix $V$
[in ${\L}^{cc}_{\mbox{\scriptsize{Y.}}}$ of Eq.~(\ref{ccpart})]
which influence the parameters $\zeta_j$.
The Cabbibo angle is fairly well determined
($\sin \theta_{12} = 0.221 \pm 0.003$), so we use the middle value
for it. On the other hand, the other two CKM rotation angles $\theta_{23}$
and $\theta_{13}$ (in the convention of Chau and Keung, or Maiani) have
the following uncertainties~\cite{particle} in the 90 percent
confidence limit
\begin{equation}
 0.032 \ < \ \sin \theta_{23} \ < \ 0.048 \ , \qquad
 0.002 \ < \ \sin \theta_{13} \ < \ 0.005 \ .
\label{ckm1}
\end{equation}
All three rotation angles lie in the first quadrant. Furthermore,
the CLEO~\cite{cleo} and ARGUS~\cite{argus} collaborations
have measured $b \to u$ transitions in semileptonic $B$ decays, with
the further resulting restriction
\begin{equation}
\frac{\sin \theta_{13}}{\sin \theta_{23}} =
{\Big |} \frac{V_{ub}}{V_{cb}} {\Big |} = 0.08 \pm 0.02 \ .
\label{ckm2}
\end{equation}
On the hand, the fourth angle $\delta$ in the CKM matrix, responsible
for CP violation, is completely undetermined yet. We note that the
restrictions (\ref{ckm1}) and (\ref{ckm2}) are mostly obtained from
weak semileptonic decays of relevant quarks and from the requirement
of unitarity. E.g., the restrictions on $|V_{cb}|$ ($\simeq
\sin \theta_{23}$) are obtained largely from ($b \to c W^- \to
c \ell^- \bar{\nu}_{\ell}$) decay; the restrictions on $|V_{ub}|$
($= \sin \theta_{13}$) are obtained largely from the requirement
of unitarity in the first row of the CKM matrix ($|V_{ub}|^2 = 1 -
|V_{ud}|^2 -|V_{us}|^2$), the two other elements of this row
($|V_{ud}|$, $|V_{us}|$) being determined largely by the comparison
of the nuclear beta decay to muon decay, and by semileptonic
decay of $K$ mesons ($K^+ \to \pi^0 e^+ \nu_e$, $K^0_L \to
\pi^{\pm} e^{\mp} \nu_e$), respectively. Also the ratio (\ref{ckm2})
is derived from semileptonic $B$ decays. Since all these decays
practically don't involve the Higgs sector (the couplings of the
Higgs to leptons are in general of the order of lepton masses,
i.e., negligible), we can argue that the restrictions (\ref{ckm1})
and (\ref{ckm2}) apply not just to the MSM, but to a large class of
standard models with extended Higgs sectors, including the present
2HDSM framework.

We performed the analysis of the $K^0$-$\bar{K^0}$ and
$B_d^0$-${\bar{B_d^0}}$ mixing according to the formulas presented
above, taking into account the uncertainties in the
experimental data for $\triangle M_{B_H^0-B_L^0}$
(\ref{deltamb}) and the hadronic uncertainties (\ref{bkgen}),
(\ref{fbprgen}) [or (\ref{bkrestr})]. The two relative
uncertainties in $\triangle M_{B_H^0-B_L^0}$ and $F_B \sqrt{B_B}$
in (\ref{mbrel})
are regarded as independent and are combined by taking the square
root of the sum of their squares. In order to account for
the uncertainty in the CKM parameters $\zeta_j$'s, we
scanned the allowed CKM parameter region (\ref{ckm1})-(\ref{ckm2})
in the $\sin \theta_{13}$ vs $\sin \theta_{23}$ plane with 169 points
($13 \times 13$) that are uniformly distributed over that region
in each direction. We allowed the CP-violating phase~\footnote{
In the {\it Review of Particle Properties}~\cite{particle} this angle
is denoted as $\delta_{13}$.}
$\delta$ in the CKM matrix $V$ to
be free, and took
for the top quark mass $m_t^{\mbox{\scriptsize{phy}}}=175 GeV$, as motivated by
the
CDF results~\cite{cdf}. The mass $m_t$ taken in the loop-integrands
(\ref{Jqcd}) and
(\ref{Iqcd}) in parameters $x_3$ ($=m^2_t/M^2_W$) and $y_3$
($=m^2_t/M^2_{H^-}$) was not the pole mass $m_t^{\mbox{\scriptsize{phy}}}$,
but the running mass $m_t(m_t) \approx 167 GeV$ (this point was
discussed in Ref.~\cite{buras}).
Then, choosing a specific value
for $M_{H^{\pm}}$, we obtain the allowed intervals of values of the
normalized Yukawa coupling $|X^{(U)}(m_t)|$
$=$$\left[ |U_{33}(m_t)| v/(\sqrt{2} m_t(m_t)) \right]$
for a chosen CKM angle $\delta$. We depicted these allowed regions
in the plane $\delta$ vs $|X^{(U)}(m_t)|$ in Figs.~3a, 3b, 3c, for the
values of the charged Higgs mass $M_{H^{\pm}} = 200, 400, 600 \ GeV$,
respectively. The solid and dash-dotted lines (and the $\delta$-axis)
delimit the allowed region for the case of the more restricted bounds
(\ref{bkrestr}) on the hadronic matrix elements that are favored by
the QCD sum rules and lattice calculations;
the dashed and dash-dot-dotted lines (and the $\delta$-axis) delimit
the allowed region for the case of less restrictive bounds
(\ref{bkgen}) and (\ref{fbprgen}). The somewhat bumpy solid lines
in these figures are a consequence of the fact that we scanned the allowed
region of the CKM parameter plane $\sin \theta_{13}$ vs $\sin \theta_{23}$
with 169 points only; increasing the number of points further would lead
to more continuous slopes of the border lines and would increase the
solid and the
dashed lines by at most a few percent, as our numerics suggests.
{}From these figures we see the following behavior. The Yukawa coupling
$|X^{(U)}(m_t)|$ has in the case of a lighter charged Higgs a more
stringent upper bound; this upper bound remains always safely
in the region ${\cal{O}}(1)$; however, it depends rather strongly on
the chosen bounds for the hadronic matrix element parameters $B_K$
and $F_B \sqrt{B_B}$ (\ref{bkgen})-(\ref{bkrestr}).
Also the
allowed region for the CP-violating CKM angle $\delta$ depends
rather strongly on these allowed values. It is the
$B_d^0$-${\bar{B_d^0}}$ mixing that restricts the angle $\delta$
from above. Furthermore, the larger the Yukawa coupling $|X^{(U)}|$ is,
the smaller is the allowed interval for $\delta$.
If $m_t^{\mbox{\scriptsize{phy}}}$ is increased, the upper bound on $|X^{(U)}|$
decreases somewhat.
In the special 2HDSM(II) and (I), the vertical axis $|X^{(U)}(m_t)|$
is to be replaced by $\cot \beta$, according to
(\ref{typeI}) and (\ref{typeII}). In this case, Figs.~3
would give us lower bounds for $\tan \beta$. Furthermore, once
the experimental and theoretical uncertainties are reduced
sufficiently, the lower bounds on $|X^{(U)}(m_t)|$ (the dash-dotted
lines in Figs.~3a-c) may increase in such a way that they don't cut the
x-axis (the line $|X^{(U)}(m_t)|=0$), i.e., the MSM would become
ruled out in such a case and we would obtain a nonzero lower bound
on $|X^{(U)}(m_t)|$.

\vspace{0.5cm}

\noindent{\large \bf {4.) $D$-$\bar D$ mixing in a general 2HDSM framework}}

\vspace{0.3cm}

For the phenomenon of the $D^0$-$\bar{D^0}$ mixing, the short distance
contribution at one loop is also represented by the diagrams of
Fig.~1. However, now the external quark legs are of the up-type
($c \bar{u}$, $\bar{c} u$), and therefore the inner quark
propagators are of the down-type. In the MSM, it turns out that the
effect of the mass of external $c$-quark, when combined with the
GIM mechanism, influences crucially the strength of the resulting
four-fermion couplings~\cite{datta2}. A relation analogous to the
$B_d^0$-${\bar{B_d^0}}$ relation (\ref{mbrel}) results in the MSM
prediction for the short distance contribution to
$\triangle M_{D^0}$ ($=|m_{D_1^0}-m_{D_2^0}|$):
$\triangle M_{D^0} \simeq 5 \cdot 10^{-18} GeV$ (when taking $F_D \simeq
0.2 \ GeV$
and $B_D \simeq 1$)~\cite{hewett}. However, long distance contributions,
e.g.~$\pi \pi$ and $K K$ intermediate states, give somewhat higher
contributions to $\triangle M_{D^0}$: the intermediate particle dispersive
approach~\cite{donoghue} gives $\triangle M_{D^0} \simeq 10^{-16} GeV$,
and the heavy quark effective theory approach~\cite{georgi} gives
$\triangle M_{D^0} \simeq 10^{-17} GeV$. On the other hand, the present
experimental upper bound for $\triangle M_{D^0}$
($\triangle M_{D^0} < 1.3 \cdot 10^{-13} GeV$) is still
crude~\cite{particle} and far above these values.
Experiments with large expected numbers
($\sim 10^8$) of reconstructed charm mesons are being planned. These
could probe whether the present upper bound should be decreased by
a factor $10^{-1}-10^{-2}$, i.e., whether $\triangle M_{D^0}
\stackrel{<}{\sim} 10^{-15} GeV$ or not. This would make the
$D$-physics more interesting, giving us possible experimental
evidence for physics beyond the MSM.

Within the present framework, the $HH$ box diagram of Fig.~1
with two $b$-quark propagators is the dominant contribution
to $\triangle M_{D^0}$, as long as the Yukawa coupling $D_{33}$
($\to X^{(D)}$) is large enough for ${\A}_{HH}^D$ [$\propto
(D_{33})^4$] to dominate over both the MSM short distance amplitude
${\A}_{WW}^D$ and over the mixed amplitude ${\A}_{HW}^D$
[$\propto (D_{33})^2$]. Therefore, we will demand that the contribution
of ${\A}_{HH}^D$ to $\triangle M_{D^0}$ not exceed the present
experimental upper bound $1.3 \cdot 10^{-13} GeV$.
Since the GIM mechanism, which suppresses
the MSM contributions, is not relevant
for such diagrams, and $m_b > m_c$, it is justifiable to ignore
the effects of the mass of the external $c$ quark. Calculation
analogous to that of the $K^0$-$\bar{K^0}$ and $B_d^0$-${\bar{B_d^0}}$
mixing, but with the QCD corrections ignored, yields for $D^0 \to
\bar{D^0}$
\begin{equation}
{\L}_{eff}^{HH} = {\A}_{HH}^D \left[ \overline{u(x)^a} \gamma^{\mu}
(\frac{1-\gamma_5}{2})c(x)^a \right]^2 \ ,
\end{equation}
\begin{equation}
\A_{HH}^D  =  - \frac{1}{128 \pi^2 M_{H^{\pm}}^2}
(\zeta_b)^2 \left[ D_{33}(m_b) \right]^4 {\I} \left( y_b,y_b \right) \ .
\label{adhh}
\end{equation}
Here, we denoted $\zeta_b = V_{cb}^{\ast} V_{ub}$ and $y_b =
(m_b/M_{H^{\pm}})^2$.
The integral for ${\I}$ is written in (\ref{inami}), and has the value
${\I}(y_b,y_b) = 1 + {\cal{O}}(y_b \ln y_b)$. Denoting by $F_D$ and
$B_D$ the hadronic matrix element parameters, analogously
as in the case of the $B_d^0$-${\bar{B_d^0}}$ mixing (\ref{fbprgen}),
we get the condition
\begin{equation}
\left( \triangle M^{HH}_D = \right) \
{\Big |} {\A}_{HH}^D {\Big |} \frac{2}{3} F_D^2 B_D M_D \ < \
1.3 \cdot 10^{-13} GeV \ ,
\end{equation}
which in turn gives us the upper bound on the Yukawa coupling
$D_{33}(m_b)$ [$\to X^{(D)}(m_b)$, cf.~(\ref{defp})]
\begin{equation}
| X^{(D)}(m_b) | \left( = \frac{|D_{33}(m_b)| v}{\sqrt{2} m_b(m_b)} \right)
\ \stackrel{<}{\sim} 26 \sqrt{\frac{M_{H^{\pm}}}{GeV}} (1 \pm 0.28) \ .
\label{dub}
\end{equation}
In this upper bound, $(\pm 0.28)$ denotes the uncertainty arising from
the experimental uncertainty in the CKM parameter $\zeta_b$ (47 percent)
and from the theoretical uncertainty in $F_D$ [we took: $F_D=0.2 \ GeV
(1 \pm 0.3)$, $B_D \simeq 1$]. Taking on the right-hand side of (\ref{dub})
the central value, we obtain for the case of
$M_{H^{\pm}}=200 \ GeV$ the upper bound $|X^{(D)}(m_b)|^{UB} \approx 362$,
which is by factor $3$ above the perturbative limit (\ref{pert}).
On the other hand, if assuming that future experiments decreased
the upper bound $(\triangle M_{D^0})^{UB}$ by two orders of
magnitude [to ${\cal{O}}(10^{-15} GeV)$], the present framework
would give as a result $|X^{(D)}(m_b)|
\stackrel{<}{\sim} 8 \sqrt{M_{H^{\pm}}/GeV}$.
For the case of $M_{H^{\pm}}=200 \ GeV$ this would imply $|X^{(D)}(m_b)|
\stackrel{<}{\sim} 110$, which is roughly at the perturbative
limit (\ref{pert}).

In order to estimate the most stringent possible constraints
imposed on $X^{(D)}$ by any future measurements of $\triangle M_{D^0}$,
we argue in the following way.
If we assume that at some future time the value of $\triangle M_{D^0}$
is well measured and turns out to be sufficiently above the value
estimated by the MSM, then the discussed general 2HDSM
framework will still remain a viable framework, in which the $HW$ and $HH$
exchange diagrams of Fig.~1 (with $b$-quark propagators) are responsible
for the deviation from the MSM prediction. The $HH$ diagrams would then
give a contribution $(\triangle M_{D^0})^{HH}$ of an order of magnitude
at least as large as the order of magnitude of the estimated MSM
contributions ($10^{-17} GeV$). Here we assume that the computed latter
contributions (the long distance contributions) will remain uncertain by
a factor of ${\cal{O}}(1)$, say a factor 2-3. This would imply
that the 2HDSM's can be successfilly distinguished from the MSM by
the measurement of $|\triangle M_{D^0}|$ only if
\begin{displaymath}
|(\triangle M_{D^0})^{HH}| \ \stackrel{>}{\sim} \ 10^{-17}GeV
\ \Longrightarrow  |X^{(D)}(m_b)| \ \stackrel{>}{\sim} \
2.4 \sqrt{M_{H^{\pm}}/GeV} \ .
\end{displaymath}
This is then approximately the lowest possible value of
$|X^{(D)}(m_b)|$ that can be inferred from any future measurements
of $|\triangle M_{D^0}|$.
Here, we took the central value $1.4 \cdot 10^{-4}$ for $\zeta_b$,
and $F_D=0.2 \ GeV$.
Assuming $M_{H^{\pm}} \stackrel{>}{\sim} 200 \ GeV$, this would
give $|X^{(D)}(m_b)|^{\mbox{\scriptsize{lowest}}} \simeq 34$, well below the
perturbative limit [Eq.~(\ref{pert})]. In the special case of the
2HDSM(II), this would imply $|X^{(U)}|^{\mbox{\scriptsize{highest}}}$
($= \cot \beta$) $\approx 0.03$. This means that within the 2HDSM(II) any
future measurements of $\triangle M_{D^0}$ would provide us with
an estimated value for $\cot \beta$ only if $\cot \beta < 0.03$
($\tan \beta > 35$). If $\tan \beta < 35$, these measurements would
not be able to distinguish between the MSM and 2HDSM(II), and
would only give an upper bound $(\tan \beta)^{UB} \simeq 35$ that
would result in $HH$ contributions comparable to the theoretical
uncertainties of the MSM long distance contributions to
$\triangle M_{D^0}$.

{}From these considerations we conclude that the future (low energy)
measurements of $\triangle M_{D^0}$ will become important for identifying a
possible 2HDSM physics beyond the MSM and for estimating the Yukawa
parameter $|X^{(D)}(m_b)|$ only in the case that the actual value of the latter
parameter is large enough: $|X^{(D)}(m_b)| \stackrel{>}{\sim} 2.4 \cdot
\sqrt{M_{H^{\pm}}/GeV}$ (for 2HDSM(II): if $\tan \beta
\stackrel{>}{\sim} 2.4 \sqrt{M_{H^{\pm}}/GeV}$),
provided that the theoretical prediction
of the long distance MSM contributions remains uncertain by a factor
${\cal{O}}(1)$. If $|X^{(D)}(m_b)|$ is smaller that this value,
the measurements of $\triangle M_{D^0}$ would not be able to
distinguish between the 2HDSM's and the MSM, and would only provide
us with the upper bound $|X^{(D)}(m_b)|^{UB} \simeq$ $2.4
\sqrt{M_{H^{\pm}}/GeV}$.

At this point, we are able to estimate the upper bound on some
of the (small) off-diagonal FCNC elements of the renormalized
Yukawa matrices $U^{(j)}$ and $D^{(j)}$ ($j=1,2$) of the starting
Lagrangian (\ref{2HD3}) (in the mass basis). As stated in the
Introduction, in the present framework we assume that these FCN Yukawa
couplings are sufficiently suppressed, so that the calculated
one-loop box-diagrams (cf.~Fig.~1) with charged Higgs
give contributions to the considered meson-antimeson mixings
which dominate over the direct tree level contributions. It was
this requirement that allowed us to ignore the off-diagonal
elements in $U^{(j)}$ and $D^{(j)}$ and to set them formally
equal to zero.
The tree level contributions of such (small) FCN Yukawa couplings
to the meson-antimeson mixings represent an exchange of a neutral
Higgs $H^0$ between the corresponding initial $q_k \bar{q_{\ell}}$
and the final $\bar{q_k} q_{\ell}$ state ($k \not= \ell$).
It is straightforward to
check that the resulting electroweak amplitudes in front of the effective
four-fermion terms in the case of the $K^0$-$\bar{K^0}$ mixing
are ${\A}^K_{\mbox{\scriptsize{tree}}} \sim $$( D_{12}^{(j)} )^2/(2 M_{H^0}^2)$
($j=1,2)$ and that they are {\it real} (if the phase between the VEV's
is $\xi = 0$). Also, the corresponding hadronic matrix elements of the
four-fermion terms are real.
Therefore, these amplitudes cannot affect the presented
analysis of the $K^0$-$\bar{K^0}$ mixing, because the relevant
quantity there was $\mbox{Im}({\A^K})$. These tree amplitudes
are of the order of $( D_{ab}^{(j)} )^2/(2 M_{H^0}^2)$ and
$( U_{cd}^{(j)} )^2/(2 M_{H^0}^2)$ for the $B^0_d$-$\bar{B^0_d}$
and $D^0$-$\bar{D^0}$ mixing, respectively, where the indices are:
$(ab)=(13)$,$(31)$ and $(cd)=(12)$,$(21)$, and $j=1,2$.
On the other hand,
the corresponding box diagrams (cf.~Fig.~1) resulted in the
amplitudes ${\A}^B_{HH} \sim$
$(\zeta^B_t)^2 {\I}(y_t,y_t) |U_{33}|^4/(128 \pi^2 M_{H^{\pm}}^2)$
$\sim 10^{-3} (\zeta^B_t)^2 |U_{33}|^4/M^2_{H^-}$,
and ${\A}^D_{HH} \sim $$10^{-3} (\zeta^D_b)^2 |D_{33}|^4/M^2_{H^-}$.
Assuming that $M_{H^0} \sim M_{H^{\pm}}$, and taking $U_{33} \sim
U_{33}^{(k)}$, $D_{33} \sim D_{33}^{(k)}$ [cf.~Eq.~(\ref{Yukawas})],
we than arrive at the following upper bounds for off-diagonal elements
of the Yukawa matrices $U^{(j)}$ and $D^{(j)}$:
\begin{equation}
\frac{|D_{13}^{(j)}|}{|U_{33}^{(k)}|^2} \ , \
\frac{|D_{31}^{(j)}|}{|U_{33}^{(k)}|^2}
\ \stackrel{<}{\sim} \ {\cal{O}} \left( 10^{-3} \right) \ ,
\qquad
\frac{|U_{12}^{(j)}|}{|D_{33}^{(k)}|^2} \ , \
\frac{|U_{12}^{(j)}|}{|D_{33}^{(k)}|^2}
\ \stackrel{<}{\sim} \ {\cal{O}} \left( 10^{-5} \right) \ ,
\ \ (j,k=1,2) \ .
\label{offdiag}
\end{equation}
Since the requirement of perturbativity of the theory (\ref{pert})
says that $|U_{33}|,|D_{33}|$$\ \leq \ {\cal{O}}(1)$, and since
in general $U_{33} \sim U_{33}^{(k)}$ and $D_{33} \sim D_{33}^{(k)}$,
we get the bounds $|D_{13}^{(j)}|,|D_{31}^{(j)}|$
$\stackrel{<}{\sim} {\cal{O}}(10^{-3})$
and $|U_{12}^{(j)}|, |U_{21}^{(j)}|$
$\stackrel{<}{\sim} {\cal{O}}(10^{-5})$ $\ (j,k=1,2)$. These two
constraints are the conditions for the presented analysis of the
$B$-$\bar B$ and $D$-$\bar D$ mixing, respectively, to hold. Furthermore,
the analysis of the
$b \to s \gamma$ decay (in Section 5) does not require any additional
restrictions on the FCN Yukawa couplings, because this decay cannot occur
at the tree level.

\vspace{0.5cm}

\noindent{\large \bf{5.) The decay $b \to s \gamma$
in a general 2HDSM framework}}

\vspace{0.3cm}

Among the loop-induced FCNC's, the $b \to s \gamma$ decay is
especially important because its strength may strongly depend on
a possible new physics, and because it has a relatively strong rate -
most of the other FCNC processes involving photons or leptons
are suppressed relatively to $b \to s \gamma$ by an order of
$\alpha_{em}$. The long range QCD interactions are here not important
because $m_b \gg \Lambda_{QCD}$. Therefore, the following approximation
is usually used (spectator model):
\begin{equation}
\frac{\Gamma(B \to X_s \gamma)}{\Gamma(B \to X_c e \bar{\nu_e})}
\approx \frac{\Gamma(b \to s \gamma)}{\Gamma(b \to c e \bar{\nu_e})} \ .
\label{spectator}
\end{equation}
The normalization to the semilpetonic rate $\Gamma(b \to c e \bar{\nu_e})$
eliminates uncertainties of the CKM matrix element $V_{ts}$ and of the
factor $m_b^5$ in the decay width $\Gamma(b \to s \gamma)$.
The recent CLEO report~\cite{cleo2} gives the measured branching
ratio
\begin{equation}
B(b \to s \gamma) =
\frac{\Gamma(b \to s \gamma)}{\Gamma(b \to \mbox{all})} \ =
(2.32 \pm 0.67) \cdot 10^{-4} \ .
\label{brcleo}
\end{equation}
This implies, at 90 percent confidence level (CL)
\begin{equation}
B(b \to s \gamma) = (2.3 \pm 1.1) \cdot 10^{-4} \ .
\label{br90}
\end{equation}
The predicted range for the MSM
($2.0 \cdot 10^{-4} < B(b \to s \gamma) <
3.8 \cdot 10^{-4}$)~\cite{bertolini}-\cite{grinstein}
is still fairly compatible with the above measurement.

The short distance QCD effects in this decay are drastic and enhance
the rate by several factors~\cite{bertolini}-\cite{buras2}. This decay
is the only known process dominated by the two-loop contributions
(i.e., leading QCD corrections). Here we use the formula for the
branching ratio derived by the method of operator product expansion
as given in Ref.~\cite{buras2} and based on the work of~\cite{ciuchini}
\begin{equation}
B(b \to s \gamma) = \frac{|V^{\ast}_{ts} V_{tb}|^2}{|V_{cb}|^2}
\frac{6 \alpha_{em}}{\pi g(z)} {\big |} C_7^{(0)eff}(\mu) {\big |}^2 \ ,
\label{btheor1}
\end{equation}
where
\begin{equation}
C_7^{(0)eff}(\mu) = \eta^{16/23} C_7^{(0)}(M_W) + \frac{8}{3}
\left( \eta^{14/23}-\eta^{16/23} \right) C_8^{(0)}(M_W) +
C_2^{(0)}(M_W) \sum_{j=1}^8 h_j \eta^{a_j} \ .
\label{btheor2}
\end{equation}
We denoted $z = m_c^{\mbox{\scriptsize{phy}}}/m_b^{\mbox{\scriptsize{phy}}}$
($\simeq 0.32$),
$\eta= \alpha_s(M^2_W)/\alpha(\mu^2)$, and
\begin{equation}
g(z) = 1 - 8 z^2 + 8 z^6 - z^8 - 24 z^4 \ln z \ .
\label{phasesp}
\end{equation}
The function $g(z)$ is the phase space factor for the semileptonic
decay $b \to c e \bar{\nu_e}$. The numbers $h_j$ and $a_j$ are given
in Ref.~\cite{buras2} and are based on Ref.~\cite{ciuchini}.
We mention that the sum of $h_j$'s is zero.
$C_j^{(0)}(M_W)$ are Wilson coefficients at some high scale [taken
to be $M_W$ ($\sim m_t \sim M_{H^{\pm}}$)] where the QCD corrections are
assumed to be negligible~\footnote{
This scale $M_W$ may be somewhat too low, especially
if $M_{H^{\pm}}$ is high: $M_{H^{\pm}} > 200 GeV$.}
\begin{equation}
C_2^{(0)}(M_W) = 1 \ ,
\qquad
C_j^{(0)}(M_W) = C_j^{(0)W}(M_W)+ C_j^{(0)H}(M_W) \ , \qquad (j=7,8) \ ,
\label{wilson0}
\end{equation}
\begin{eqnarray}
C_7^{(0)W}(M_W)&=& \frac{3 x^3-2x^2}{4 (x-1)^4} \ln x +
 \frac{-8 x^3-5 x^2 + 7 x}{24 (x-1)^3} \ ,
\nonumber\\
C_8^{(0)W}(M_W)& =& \frac{- 3 x^2 }{4 (x-1)^4} \ln x +
 \frac{-x^3 + 5 x^2 + 2 x}{8 (x-1)^3} \ ,
\label{wilson1}
\end{eqnarray}
\begin{eqnarray}
C_7^{(0)H}(M_W)& =& X^{(U)} X^{(D)} \left[ \frac{3 y^2 - 2 y}{6 (y-1)^3}
 \ln y + \frac{-5 y^2 + 3 y}{12 (y-1)^2} \right] +
\nonumber\\
& & +
\left( X^{(U)} \right)^2 \left[ \frac{3 y^3 - 2 y^2}{12 (y-1)^4}
\ln y + \frac{-8 y^3 - 5 y^2 + 7 y}{72 (y-1)^3} \right],
\nonumber\\
C_8^{(0)H}(M_W)& =& X^{(U)} X^{(D)} \left[ \frac{- y}{2 (y-1)^3}
 \ln y + \frac{- y^2 + 3 y}{4 (y-1)^2} \right] +
\nonumber\\
& & +
\left( X^{(U)} \right)^2 \left[ \frac{- y^2}{4 (y-1)^4}
\ln y + \frac{- y^3 + 5 y^2 + 2 y}{24 (y-1)^3} \right].
\label{wilson}
\end{eqnarray}
We denoted $x = (m_t/M_W)^2$ and $y = (m_t/M_{H^{\pm}})^2$.
$C_7^{(0)W}$ and $C_7^{(0)H}$ are Wilson coefficients associated with the
leading one-loop electroweak diagrams generating the $b \to s \gamma$
transition (Figs.~4a, b, respectively), i.e., these diagrams induce the
term
\begin{equation}
{\L}_{eff}^{ew} = \frac{G_F}{\sqrt{2}} V_{ts}^{\ast} V_{tb}
C_7^{(0)}(M_W) {\hat {\cal{O}}_7} \ ,
\qquad
{\hat {\cal{O}}_7} = \frac{e_0}{8 \pi^2} m_b {\bar{s_a}}
\sigma^{\mu \nu} (1 + \gamma_5) b^a F_{\mu \nu} \ ,
\label{o7}
\end{equation}
where $\sigma^{\mu \nu} = \frac{i}{2} [\gamma^{\mu}, \gamma^{\nu}]$
and $F_{\mu \nu}$ is the usual QED field strength tensor. Note that
$C_7^{(0)eff}(\mu = M_W) = C_7^{(0)}(M_W)$, i.e., this is the contribution
when we ignore the QCD effects. Similarly, $C_8^{(0)W}(M_W)$ and
$C_8^{(0)H}(M_W)$ are Wilson coefficients corresponding to an
analogous operator, but with $F_{\mu \nu}$ replaced by the
corresponding gluon field strength tensor $G^a_{\mu \nu}$,
i.e., they can be derived from the one-loop diagrams of Figs.~4a,b,
respectively, when we
replace the photon external leg by a gluon external leg.
One new element in Eqs.~(\ref{wilson1})-(\ref{wilson}) are
the expressions for $C_j^{(0)H}$ (j=7,8), corresponding to the
diagrams of Fig.~4b. These expressions have been written in the
literature for the sprecial cases of the 2HDSM(I) and (II);
here, they are valid for the discussed general 2HDSM framework.
As we see, they involve {\it both} normalized Yukawa couplings of
the charged Higgs: $X^{(U)}(E)$ and $X^{(D)}(E)$. Therefore,
this fact will provide us with interesting constraints on the framework
in the plane $X^{(U)}$ vs $X^{(D)}$.
As already argued in Section 3, the effective energy of probes
$E$ in the couplings $X^{(U)}(E)$ and $X^{(D)}(E)$ in Eq.~(\ref{wilson})
is $E \approx m_t$, because the top
quark gives the dominant contributions
to the relevant 1-loop diagrams (cf.~Fig.~4b).

In the calculation, we take the low energy cut-off $\mu$ to be
$\mu = m_b \simeq 4.9 \ GeV$; $\alpha_s(M_W^2) \simeq 0.119$,
$\alpha_s(m_b^2) \simeq 0.22$, $|V_{ts}^{\ast} V_{tb}|^2/|V_{cb}|^2
\simeq 0.95$; $m_c^{phy}/m_b^{\mbox{\scriptsize{phy}}} \simeq 0.32$;
$\alpha_{em}(m_b^2) \simeq \alpha_{em}(1 GeV^2)
\simeq 1/137$, $B(b \to c e \bar{\mu_e}) \approx B(b \to e \bar{\mu_e}+
\mbox{anything}) \simeq 0.104$. By the same argument as in the case of the
$B$-$\bar{B}$ and $K$-$\bar{K}$ mixing, we take for $m_t(E)$
in the parameters $x$ and $y$ in Eqs.~(\ref{wilson1}) and (\ref{wilson})
the value $m_t(m_t) \simeq 167 \ GeV$, which corresponds to
$m_t^{\mbox{\scriptsize{phy}}} = 175 \ GeV$. With these input data, the MSM
prediction [i.e., when setting $C_j^{(0)H}(M_W)=0$ in (\ref{wilson0})
and (\ref{btheor2})] is: $B(b \to s \gamma)^{\mbox{MSM}} \simeq 2.90 \cdot
10^{-4}$.
As discussed in Ref.~\cite{buras2}, the major uncertainties in the formula
(\ref{btheor1})-(\ref{btheor2}) come from the $\mu$-dependence
of the QCD parameter $\eta = \alpha_s(M^2_W)/\alpha_s(\mu^2)$
($m_b/2 < \mu < 2 m_b$). This theoretical uncertainty, together with
other experimental and theoretical uncertainties, would result in
an uncertainty of up to 30 percent in this formula~\cite{buras2}. Combining
this 30 percent uncertainty with the 90 percent CL result (\ref{br90}) of
CLEO, we obtain the following limits for the allowed values of the formula
(\ref{btheor1}):\\
\centerline{$ 0.92 \cdot 10^{-4} \ <  \
B(b \to s \gamma)^{\mbox{\scriptsize{th.}}}
\ < \ 4.86 \cdot 10^{-4}$.}\\
\normalsize
The superscript ``th.'' denotes the expression
(\ref{btheor1})-(\ref{btheor2}). For chosen values $M_{H-} =$$200$,
$400$, $600 \ GeV$, these limits
in the discussed
2HDSM framework result in the allowed regions of the plane
$X^{(U)}(m_t)$ vs $X^{(D)}(m_t)$ as depicted in Figs.~5a,b,c,
respectively. For comparison, the dashed
line in these figures represents the region
of the 2HDSM(II) ($X^{(D)} = 1/X^{(U)} = \tan \beta$), and the
dash-dotted line the region of the 2HDSM(I) ($X^{(D)} = - X^{(U)} =
- \cot \beta$). Incidentally, we see from Fig.~5a that in the case
of $M_{H^{\pm}}= 200 GeV$ the line for the 2HDSM(II) lies entirely
outside the allowed region, i.e., in the ``type II'' model we must have
$M_{H^{\pm}} > 200 GeV$ as a consequence of the $b \to s \gamma$ data.
In Figs.~5a-c we included only those regions for $X^{(U)}$ which were
allowed by $B_d^0$-${\bar{B_d^0}}$ and $K^0-\bar{K^0}$ mixing for the
choice (\ref{bkrestr}) of the range of the hadronic parameters
$B_K$ and $F_B \sqrt{B_B}$ (cf.~Figs.~3a-c).

\vspace{0.5cm}

\noindent{\large \bf{6.) Conclusions}}

\vspace{0.3cm}

{}From the presented discussion we infer that the future
improved experimental
data for the $B$-$\bar B$, $K$-$\bar K$ and $D$-$\bar D$ mixing and for
the $b \to s \gamma$ decay, as well as the future reduced
theoretical uncertainties for the hadronic matrix elements,
can further severely constrain the (dominant) Yukawa couplings
$X^{(U)}$, $X^{(D)}$ of the charged Higgs.
If the experimental data are improved sufficiently and the theoretical
hadronic uncertainties reduced sufficiently, then the analyses such as
the one presented here may in a foreseeable future be able to rule out
the minimal SM and even certain special types of the 2HDSM's
(e.g., the popular ``type II'', and ``type I'' 2HDSM). In such a case,
a more general 2HDSM framework discussed here could still survive as
a viable framework.
We note that this can be achieved by analysing {\it low} energy
physics alone - the meson-antimeson mixings and the $b$ decays
are phenomena at low energy.
If the charged Higgs is found in high energy experiments
and its mass measured, this mass
will represent important new information. This information would
greatly facilitate the analyses of low energy phenomena
and would help to either rule out or to favor various specific
2HDSM scenarios that are contained in the discussed 2HDSM framework.

\vspace{0.6cm}

\newpage

\noindent{\large \bf{Acknowledgments}}

\vspace{0.2cm}

The author would like to thank to E.A.~Paschos for bringing to his attention
the question of general 2HDSM frameworks with suppressed FCNC's, and
to P.~Overmann and A.~Datta for helpful and stimulating discussions.


\newpage

\small \normalsize

\noindent
\noindent{\normalsize \bf{Figure Captions}}

\vspace{1.cm}

\noindent
{\small {\bf Fig.~1}: Box diagrams giving short distance
contributions to meson-antimeson mixing ($P^0$= $B^0$, $K^0$, $D^0$);
solid lines denote quarks.}

\vspace{1cm}

\noindent {\small {\bf Figs.~2a-d}: Various possible leading-log
QCD corrections to the box diagrams; dash-dotted line
denotes a gluon propagator.}

\vspace{1cm}

\noindent {\small {\bf Figs.~3a, 3b, 3c}: The allowed regions in the plane
$\delta$ vs $|X^{(U)}|$, for the values of the charged
Higgs mass $M_{H^{\pm}} = 200, 400, 600 \ GeV$, respectively.
The solid and dash-dotted lines (and the $\delta$-axis)
delimit the allowed region for the case of the more restricted bounds
(\ref{bkrestr}) on the hadronic matrix elements;
the dashed and dash-dot-dotted lines (and the $\delta$-axis) delimit
the allowed region for the case of less restrictive bounds
(\ref{bkgen}) and (\ref{fbprgen}).}

\vspace{1.cm}

\noindent
{\small {\bf Figs.~4a-b}: The dominant one-loop contributions to the
$b \to s \gamma$ decay. Besides the top quark propagator, the loops
contain physical $W^{\pm}$ (Fig.~4a) and physical $H^{\pm}$ (Fig.~4b).}

\vspace{1.cm}

\noindent
{\small {\bf Figs.~5a-c}: The regions in the $X^{(U)}(m_t)$ vs
$X^{(D)}(m_t)$ plane that are allowed by the data on the
$b \to s \gamma$ decay, for $M_{H^{\pm}}= 200,400,600 \ GeV$,
respectively. The allowed region is the central region between the
four solid curves near the axes, and two separate stripes between the
solid lines in the upper left and the lower right part of each graph.
The dashed line corresponds to the
special ``type II'' 2HDSM ($X^{(D)}(m_t)=$$1/X^{(U)}(m_t)=$$\tan \beta$),
and the dash-dotted line to the ``type I'' 2HDSM
($X^{(D)}(m_t)=$$-X^{(U)}(m_t)=$$-\cot \beta$.}

\end{document}